\documentclass[aps,prl,numerical,reprint,superscriptaddress,showpacs]{revtex4-1}
\usepackage{color}
\usepackage[version=3]{mhchem} % Formula subscripts using \ce{}
\usepackage{gensymb}
\usepackage{amssymb}
\usepackage{upgreek}
\usepackage{graphicx}
\usepackage{amsmath}
\usepackage[normalem]{ulem} % for strikeout; [normalem] prevents unintended changes in bibliography formatting
\usepackage[usenames,dvipsnames]{xcolor}
\usepackage{xr}
\newcommand{\uvec}[1]{\boldsymbol{\hat{\textbf{#1}}}}

\begin{document}

\title{Emergent pattern formation in an interstitial biofilm}

\author{Cameron Zachreson}
\affiliation{School of Mathematical and Physical Sciences, University of Technology Sydney, Ultimo, New South Wales 2007, Australia}
\author{Christian Wolff} 
\affiliation{School of Mathematical and Physical Sciences, University of Technology Sydney, Ultimo, New South Wales 2007, Australia}
\author{Cynthia Whitchurch}
\affiliation{The ithree institute, University of Technology Sydney, Ultimo, New South Wales 2007, Australia}
\author{Milos Toth}
\affiliation{School of Mathematical and Physical Sciences, University of Technology Sydney, Ultimo, New South Wales 2007, Australia}

\begin{abstract}
Collective behavior of bacterial colonies plays critical roles in adaptability, survivability, biofilm expansion and infection. We employ an individual-based model of an interstitial biofilm to study emergent pattern formation based on the assumptions that rod-shaped bacteria furrow through a viscous environment, and excrete extracellular polymeric substances which bias their rate of motion. Because the bacteria furrow through their environment, the substratum stiffness is a key control parameter behind the formation of distinct morphological patterns. By systematically varying this property (which we quantify with a stiffness coefficient $\gamma$), we show that subtle changes in the substratum stiffness can give rise to a stable state characterized by a high degree of local order and long-range pattern formation. The ordered state exhibits characteristics typically associated with bacterial fitness advantages, even though it is induced by changes in environmental conditions rather than changes in biological parameters. Our findings are applicable to broad range of biofilms and provide insights into the relationship between bacterial movement and their environment, and basic mechanisms behind self-organization of biophysical systems.

\end{abstract}

\date{\today}

\maketitle

\section{introduction}
%emergence to models to infections to detail
Emergent phenomena and collective motion of bacteria play important roles in the proliferation of bacterial biofilms (dense colonies of surface-associated bacteria), and may produce significant survival advantages in diverse environments. The fundamental principles of collective motion have been explored extensively using minimal simulation models of moving, interacting individuals \cite{vicsek2012collective,vicsek1995novel,ginelli2010large}, and the prospect of applying the insights gained from such studies to infection and disease is very attractive. However, most minimal models of collective motion are not directly applicable to bacterial colonies due to oversimplification of the interactions between individuals, and between individuals and their enclosing, physical environment. 

The gap between the real world of bacteria and idealized, minimal simulations has been bridged to some extent by system-specific models based on physical interactions between motile bacteria \cite{peruani2012collective,peruani2006nonequilibrium,balagam2015mechanism}. In parallel, biophysical models of non-motile bacteria have shown how the physical interactions between growing cells and the soft material surrounding them can determine group morphology \cite{grant2014role,ghosh2015mechanically,farrell2013mechanically}. 

Here we investigate collective behavior of an interstitial biofilm using an individual-based model that accounts for stigmergy (ie: path following) \cite{Heylighen201650,Heylighen20164,gloag2013self,gloag2013stigmergy} by bacteria that comprise the biofilm. We show that subtle differences in environmental conditions can cause dramatic shifts in the emergent, collective behavior of the colony due to stigmergic effects. Specifically, an increase in substratum stiffness yields a transition from dispersed, disordered movement, to a highly stable, structured morphology characterized by a high local density of constituent bacteria that move efficiently within self-constructed channels. The ordered state, controlled by physical properties of the environment, results in a condition which reconciles two often mutually exclusive properties that are typically associated with fitness advantages: high mobility and high local density. Our findings provide insight into the relationship between bacterial movement mechanics and the physical properties of the enclosing environment and, more broadly, into the nature  of emergence and pattern formation in complex biophysical systems. Understanding such processes can help explain adaptation to specialized habitats, and inform the design of experiments and biomedical surfaces.

\section{model}

\begin{figure}[b!]
{\includegraphics[width=0.95\linewidth]{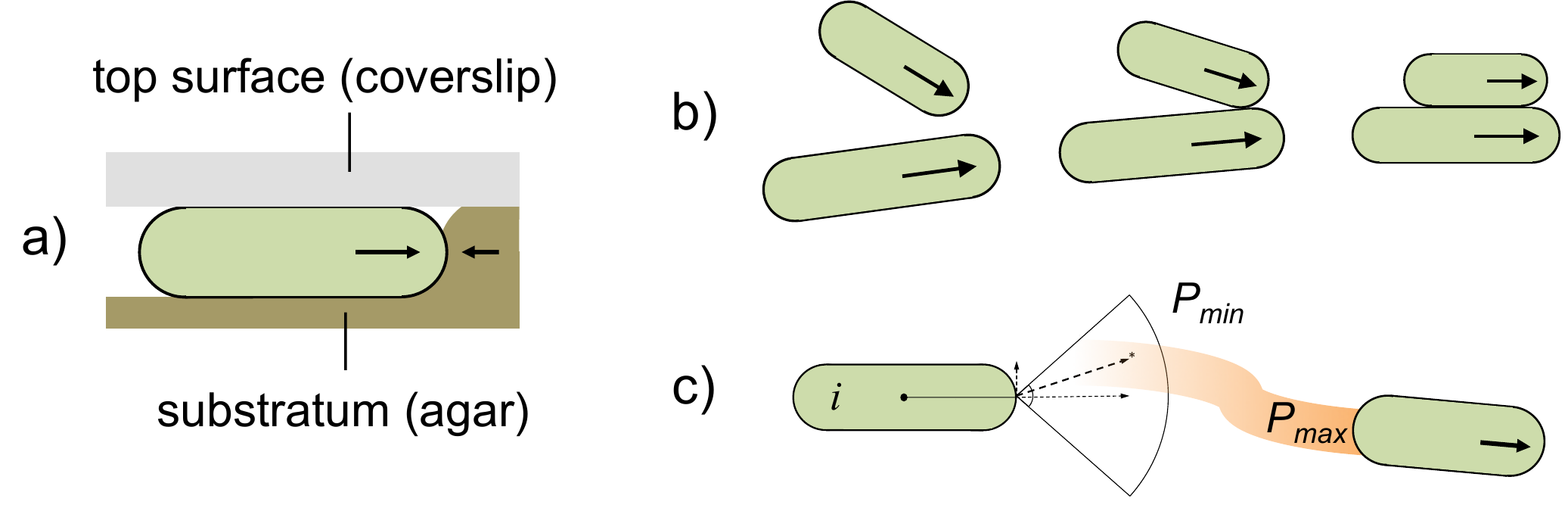}}
\caption{{Primary processes implemented in the bacterium behavioral model.} (a) Furrowing through an elastic substratum.  (b) Alignment of rod-shaped bacteria through physical repulsion. (c) Movement bias induced by excreted extracellular polymeric substances. Bacterium $i$ samples the space accessible by its pili by randomly selecting a point within an arc extending from a forward pole. The pilus attachment probability scales with EPS concentration (shown in orange), up to a maximum value $P_{max}$. In the absence of EPS, the attachment probability is given by $P_{min}$.}
\label{model}
\end{figure}

% model summary

We implemented an individual-based model of bacterial behavior in a quasi-2D environment (Fig. \ref{model}). It consists of two primary components: (i) a realistic model of physical interactions between rigid, rod-shaped bacteria confined in the space between a viscous medium and a solid surface (Fig. \ref{model}a, b), in which the resistance to motion due to deformation of the substratum is quantified by a single parameter (a stiffness coefficient, $\gamma$), and (ii) bacterial motility that is enabled by the extension and retraction of Type IV Pili (T4P) and biased by extracellular polymeric substances (EPS) which are excreted by the bacteria (Fig. \ref{model}c).  We intentionally exclude complex phenomena such as chemotaxis and contact-based signalling in order to show that explicit simulation of high order biochemical communication processes is not necessary to explain this type of biofilm self-organisation and pattern formation. Instead, morphogenesis results from two basic forms of stigmergy arising from the formation of furrows in a viscous substratum and following of EPS trails deposited by the moving bacteria. EPS following is implemented as preferential binding of pili to the EPS \cite{van2005dna,li2003extracellular}, as shown in Fig. \ref{model}c. This general phenomenon is ubiquitous to a broad range of bacteria such as {\it{P. aeruginosa}} and {\it{M. xanthus}} that use Type IV Pili for movement while actively modifying the surface properties through substance excretion \cite{van2005dna,maier2015bacteria,li2003extracellular}.

We note that in this model, the EPS does not directly cause alignment of neighboring bacteria. Instead, alignment results from collisions of the rigid, rod-shaped individuals  (Fig. \ref{model}b). We parametrized the model based on experimental data gathered for {\it{Pseudomonas aeruginosa}}. It is, however, flexible and directly applicable to a broad range of biofilms that use T4P for surface motility \cite{pelicic2008type}.

% physical properties 
\subsection{physical properties of individuals}
Each bacterium is simulated as a sphereocylindrical particle (a rod of length $l$ and width $w$, with hemispherical caps of radius $w/2$). We assume overdamped dynamics due to the high viscosity of the medium. The physical interaction scheme is a generic model of collisions between rod-shaped objects and our numerical implementation of this scheme is similar to that used in Ref. \cite{ghosh2015mechanically,farrell2013mechanically}. Friction is uniform in space, so that damping does not vary along the length of a rod-shaped particle. Translational and angular velocities that arise from physical interactions therefore depend on particle length and a constant friction coefficient $(\mu)$, and are proportional to the applied forces and torques, respectively. 

Translational motion is calculated using:
\begin{equation}
\frac{dx}{dt} = \frac{\vec{F}}{{\mu}l} \,,
\end{equation}
here, $\vec{F}$ is the sum of forces generated by motility $\vec{F}_{p}$, environmental forces $\vec{F}_{s}$ , and particle-particle interactions $\vec{F}_{ij}$ so that $\vec{F} = \vec{F}_{p} + \vec{F}_{s} + \sum_{j}\vec{F}_{ij}$.

Similarly, angular velocity is given by:
\begin{equation} 
\frac{d\theta}{dt} = \frac{12\tau}{{\mu}l^3}\,,
\end{equation}
in which $\tau = \tau_{p} + \tau_{s} + \sum_{j}\tau_{ij}$ is the net torque on particle $i$ due to motility $\tau_{p}$, environmental potentials $\tau_{s}$, and interaction forces $\tau_{ij}$.

\subsection{repulsive interaction forces}

To model the repulsive force of collisions between cell walls, we employ a potential (see methods) that yields repulsion when the distance between particles is smaller than the rod width ($d_{ij} < w$). The force $\vec{F}_{ij}$ acts  along ${\uvec{r}_{ij}}$, where $\uvec{r}_{ij}$ is a unit vector defined by the closest points between particle backbones (Fig. \ref{FigureS2}).  $\vec{F}_{ij}$ also generates torque of magnitude $\tau_{rep} = [\vec{F}_{ij\perp}][ r_{lev}]$, where $r_{lev}$ is the lever arm distance and  $ \vec{F}_{ij\perp}$ is the component of the interaction force perpendicular to the long axis of the particle. 

% movement mechanism

\subsection{twitching motility}

On solid surfaces, bacteria such as {\it Pseudomonas aeruginosa, Myxococcus xanthus}, and {\it Neisseria gonorrhoeae} can express a motility apparatus termed Type IV Pili: strands of protein that are extruded through on-demand assembly from monomeric precursors, attach to a surface, and retract via dis-assembly. The force of retraction pulls the cell forward, and repeated retractions and extensions allow cell motion across a surface \cite{skerker2001direct}. The behavior associated with this type of movement is called `twitching' motility \cite{mattick2002type}. 

\begin{figure}[h!]
\includegraphics[totalheight=0.6\textheight]{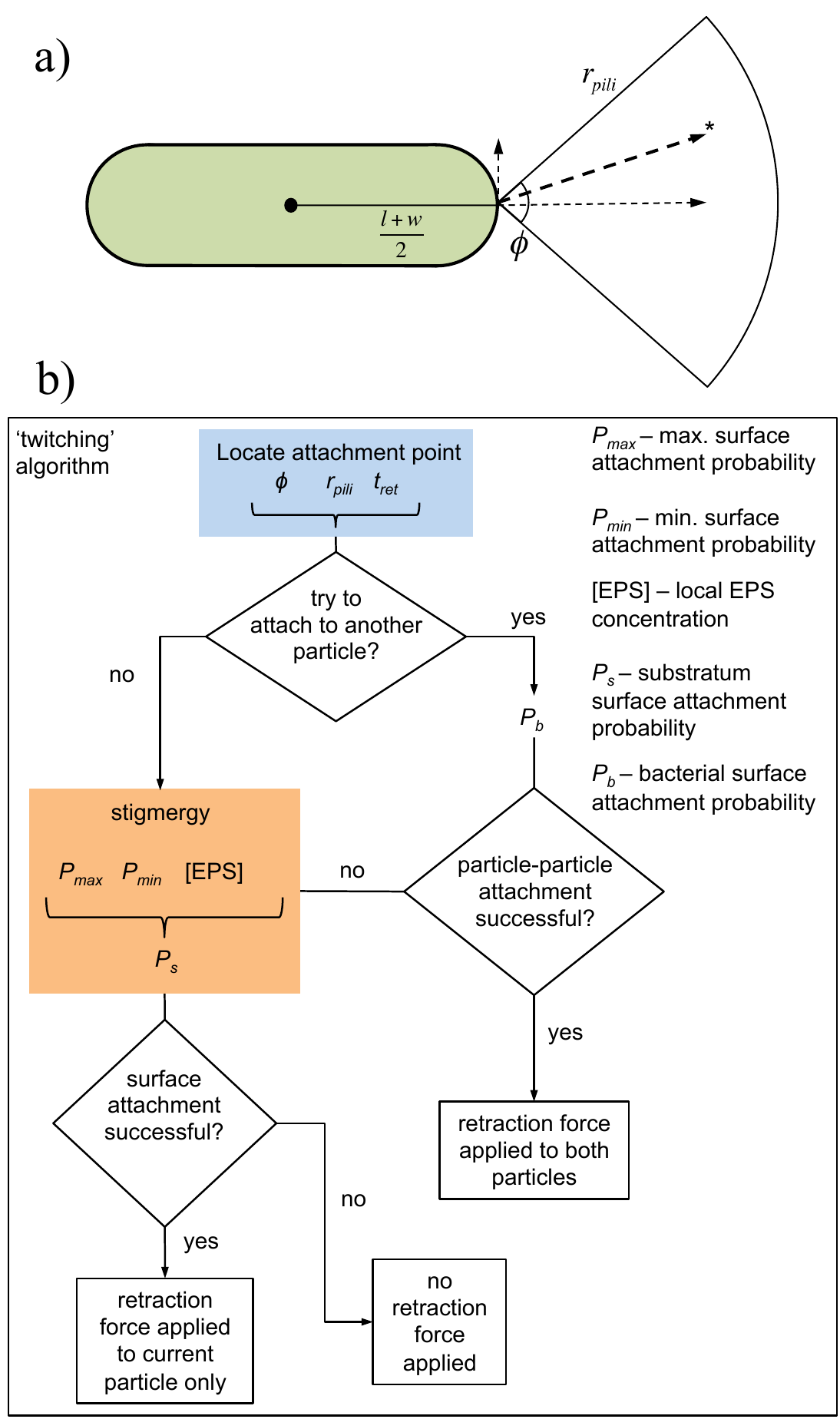}
\caption{(a) Twitching scheme: the attachment point (*) is selected at random  in the region defined by $\phi$ and $r_{pili}$. The retraction force acts along the associated vector shown as a bold dotted line. (b) Flowchart describing how the movement of a single particle is calculated from the location of a random attachment point, the analysis of the associated binding probability, and the application of the resulting force. }
\label{FigureS1}
\end{figure}

To model cell movement we specify an angle $\phi$ and a pilus range $r_{pili}$ that define an area bound by an arc extending from the front pole of a particle (Fig. \ref{FigureS1}). To produce forward motion of a particle, a point is selected at random in this region. If a binding event occurs, a force $\vec{F_{p}}$ is applied between the particle's leading pole and the binding site over the duration of the retraction time (which is selected at random from a uniform distribution $[0, 2t_{ret}]$ defined by the the mean retraction period $t_{ret} = 5s$). 

Pilus retraction along the vector between the leading pole of the particle and the binding site generates a torque of magnitude given by: 
\begin{equation}
\tau_{p} = \vec{F}_{p\perp} \frac{l + w}{2}\,,
\end{equation}
 where  $\vec{F}_{p\perp}$ is the component of the retraction force normal to the long axis of the particle. 

If the T4P binding site happens to be on the body of another particle, the binding probability is $P_{b}$ (constant throughout the simulation) and the retraction force is split between both particles, directing them towards each other. An appropriate torque is applied to the particle being pulled with the attachment point approximated as the point on the midline that is nearest the binding site. Movement of both particles is accounted for during subsequent time increments associated with the twitching event. 

The parameters $|\vec{F}_{p}|$, $r_{pili}$, $\phi$, and $t_{ret}$ control the orientational persistence time and movement velocity in the absence of interactions with other particles or any environmental forces. If the space in front of a cell is heterogeneous, the pili act as sensors as well as motility machines. In our model this is relevant to the process of EPS trail following, where the retraction duration determines the frequency at which the environment is sampled and modulates the ability for bacteria to effectively follow EPS trails. Here, we constrained this value based on experimentally determined retraction distances and movement rates (see `parameterization' section).

{\it P. aeruginosa} and other rod shaped bacteria such as {\it M. xanthus}, exhibit spontaneous reversals of polarity during twitching motility \cite{mattick2002type}. Since T4P are localized to one pole of the cell, this process reverses the motility direction of single bacteria. To implement spontaneous polarity reversal, we define a reversal period that is selected at random from a Gaussian distribution defined by the average reversal period $t_{rev}$ and its standard deviation $\sigma_{rev}$. At the moment of reversal, a countdown starts and a new reversal event occurs when this clock reaches zero.

% stigmergy
\subsection{stigmergy}

\begin{figure}[h!]
\includegraphics[width=0.45\textwidth]{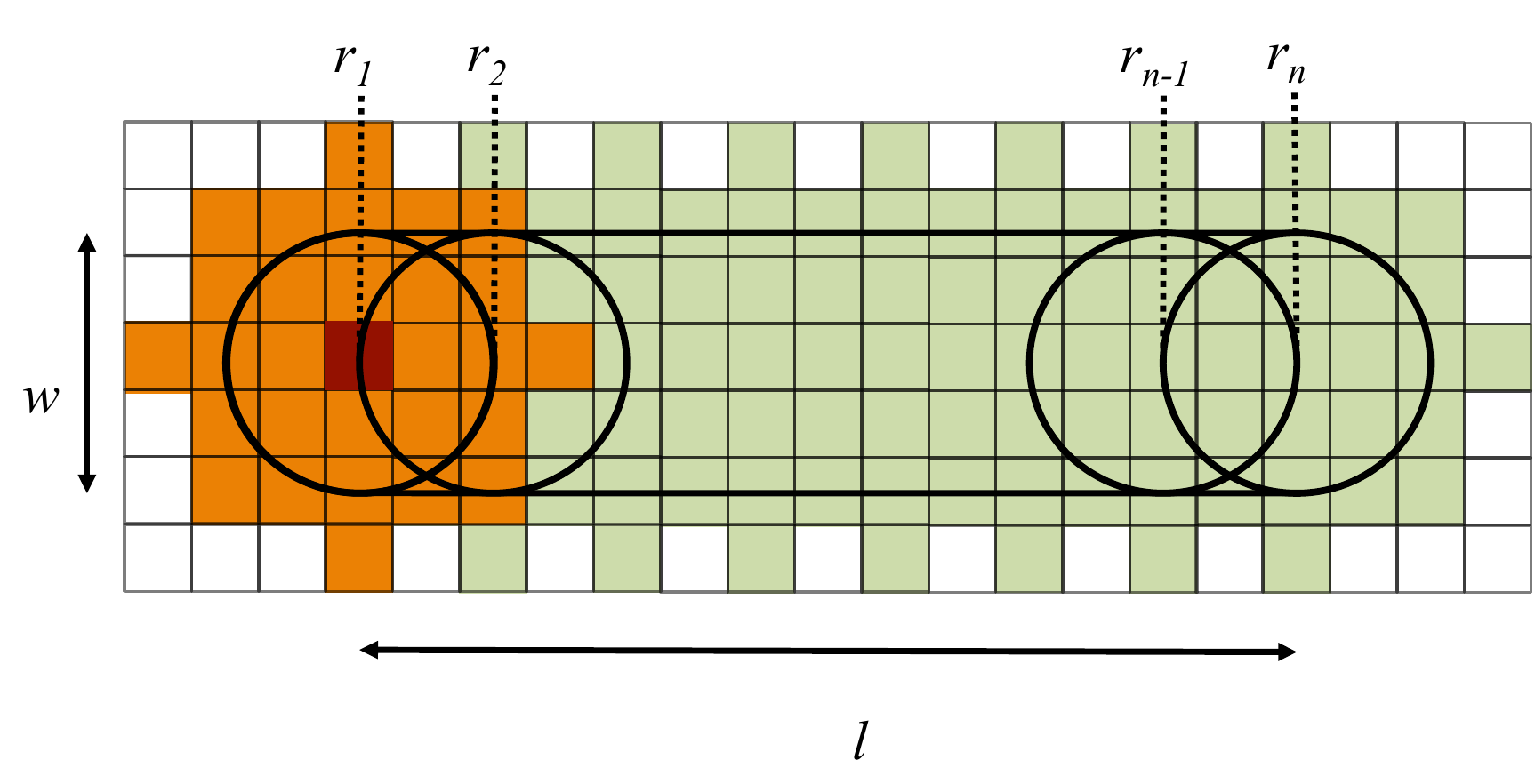}

\caption{A discretized rod in two spatial dimensions. The squares represent the pixels used for accumulation of stigmergy tracer counts. The orange squares represent those used for calculating the local forces from substratum resistance at $r_{1}$, the discrete pixel corresponding to the center of segment $1$ is colored in red.}
\label{FigureS3}
\end{figure}

The simulation takes place in finite 2D space with periodic boundary conditions. The space is discretized into a square mesh with a pixel resolution of $\Delta x = w/4$, where $w$ is the particle width of $1~\mu$m. The pixels are used to accumulate a stigmergy tracer as the particles traverse the simulation space (Fig. \ref{FigureS3}). We used Eulerian integration to calculate changes in tracer level.

Particles deposit a trace in their environment represented by counts that decay exponentially in time and accumulate when a particle is present. Count accumulation in each pixel of area $[\Delta x]^2$ saturates according to:

\begin{equation}
{\frac{dC}{dt}}^{+} = k\frac{C_{max} - C(t)}{C_{max}}[\Delta x]^2\,,
\label{countvstp}
\end{equation}

where $k$ is the trace deposition rate per unit area and $C_{max}$ is the maximum tracer count per pixel.
Accumulation competes with exponential decay taking place at rate $\beta$: 

\begin{equation}
  {\frac{dC}{dt}}^{-} = -{\beta}C(t)\,.
  \label{countvstm}
\end{equation}

Because of the competition, a steady-state value $C_{st} \leq C_{max}$ is reached if a particle moves sufficiently slowly:

\begin{equation}
C_{st} =\Big[\frac{1}{C_{max}}+\frac{\beta}{k}\Big]^{-1}\,.
\end{equation}

The steady-state value $C_{st}$ is a function of the free parameter $C_{max}$, which represents the limit to which the environment can sustain continual modification.
   
There are two types of counts, $C_{s}$ and $C_{p}$, each with their own values of $k$ and $\beta$ ( $k_{s}$, $\beta_{s}$ and $k_{p}$, $\beta_{p}$, respectively). $C_{s}$ accounts for the formation of furrows in the substratum and $C_{p}$ accounts for the effects of excreted EPS on the probability of T4P attachment to the surface.

 % furrowing
\subsection{stigmergy mechanism 1: furrowing}
The substratum-coverslip interface resists debonding, thereby generating a force that resists bacterial motion and biofilm expansion. In our model, this force is calculated by allowing a surface deformation potential $U_{s}$ to scale with the local value of $C_s$ and the stiffness coefficient $\gamma$: 

\begin{equation}
\gamma = \frac{dU_s}{dC_s} \,,
\end{equation}

\begin{equation}
\vec{F}_{s}(x, y) = \gamma \nabla C_{s}(x, y)\,,
\end{equation}

where $\nabla C_{s}(x, y)$ is the local central difference between neighboring pixels $\Delta C_s$ divided by the pixel spacing $\Delta x$. 

\begin{figure}[h!]
\includegraphics[totalheight=0.3\textheight]{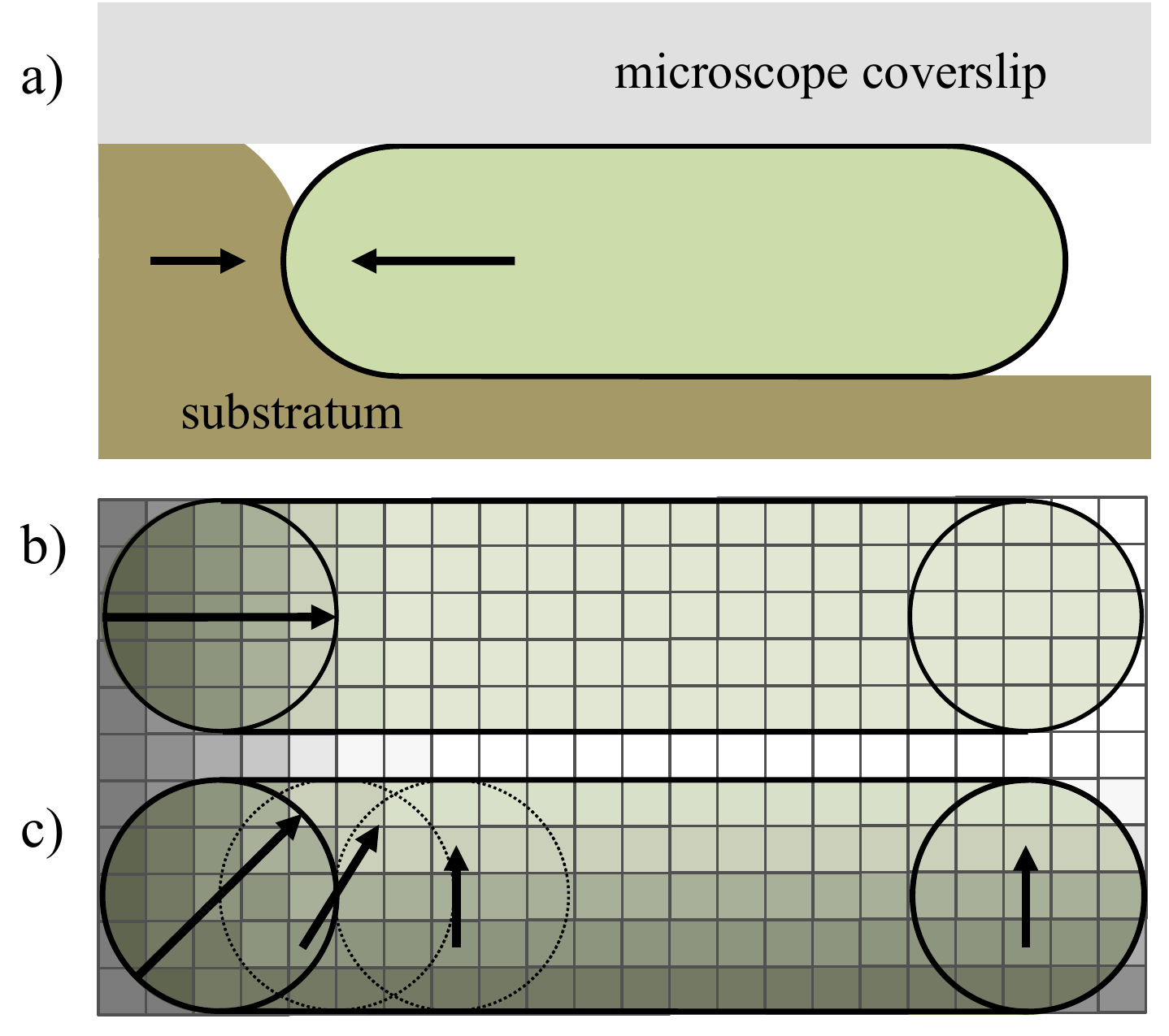}
\caption{Forces are generated due to deformation of the substratum. (a) Schematic illustration of a bacterium furrowing into the interstitial space  where it must overcome resistance from the substratum. (b, c) Top-down schematic of simulations where spatial gradients from low trace levels (dark pixels) to high (light pixels) correspond to a force field that resists particle motion due to the elastic properties of the substratum. If the gradient is aligned with the long axis of a particle, it generates a translational force only (b). A non-uniform gradient aligned with the short axis of the particle gives rise to torque as well, which is evaluated at each rod segment (c). }
\label{FigureS4}
\end{figure}

To find the force applied at each rod segment, $\vec{F}_{s}(x, y)$ is integrated over the pixels $(x, y)$ within the area each rod segment $A_{seg}$ centered at position $r$ where: 
\begin{equation}
 r\in{\{}-\frac{l}{2}, -\frac{l - w}{2},  -\frac{l - 2w}{2} ... \frac{l - w}{2}, \frac{l}{2}{\}}\,,
\end{equation}

\begin{equation}
\vec{F}_{s}(r) = \sum_{(x,y)\in A_{seg}} F_{s}(x, y)\,,
\end{equation}

\begin{equation}
\vec{F}_{s} = \sum_{r}\vec{F}_{s}(r)\,,
\end{equation}

so that the appropriate torques can be applied:

\begin{equation}
\tau_{s} = \sum_{r} \vec{F}_{s\perp} r\,.
\end{equation}
Here $\gamma$ scales the resistance felt by a moving rod with respect to the local topographical gradient (Fig. \ref{FigureS4}). Physically, $\gamma$ is related to the elastic properties of the substratum. Capillary forces are likely responsible for the observed substratum reformation in the presence (but not in the absence) of a coverslip \cite{gloag2013self}, and would be relevant to $\gamma$, $\beta_{s}$, and $k_{s}$. Because stiffness would affect the rate at which the substratum responds to the presence of a bacterium and the rate at which furrows refill due to capillary forces, we make $\beta_{s} \propto \gamma^{-1}$ and $k_{s} \propto \gamma^{-1}$, which is equivalent to stating:

\begin{equation}
\frac{dU_s}{dt} = \gamma\frac{dC_s}{dt}
\end{equation}

is independent of $\gamma$. This means we can re-cast equations \ref{countvstp} and \ref{countvstm} as:
\begin{equation}
\frac{dU_s}{dt}^{+} = k_U\frac{C_{max} - C(t)}{C_{max}}[\Delta x]^2\,,
\end{equation}
and
\begin{equation}
\frac{dU_s}{dt}^{-} = -\beta_UC(t)\,,
\end{equation}
respectively, where $k_U = \gamma k_s = 0.05$ and $\beta_U = \gamma \beta_s =  2.5 \times 10^{-4}$ are constant for all values of $\gamma$, and describe the deformation and restitution rates of the potential field corresponding to surface topography. That is, the capillary forces and deformation forces produced by the bacteria are assumed to be independent of substratum stiffness. Consequently, it takes bacteria longer to deform a stiffer substratum, but the resulting deformation will resist restitution, and persist longer in the absence of bacteria. The resitution rate must be significantly slower than the movement rate of bacteria in order for furrows to persist in the absence of bacteria. If the restitution rate is on the order of the individual movement rate, the force generated by substratum deformation resembles surface tension that holds clusters of cells together.

% EPS

\subsection{stigmergy mechanism 2: EPS excretion}

Based on literature precedent \cite{van2005dna,maier2015bacteria,li2003extracellular}, the proposed role of EPS in motility is modeled by making the local probability of pilus attachment to the substratum, $P_{s}$, scale with the the fraction $K$ of the local area covered by EPS binding sites $K = C_{p}(x, y)/[\Delta x]^2$, so that movement is biased along trails. 

\begin{equation}
P_{s}(K) = KP_{max} + (1 - K)P_{min}\,.
\end{equation}
This value is bounded from below by a finite probability $P_{min}$, which corresponds to the natural binding affinity of T4P to the unaltered surface and defines the probability of attachment when $K = 0$. Similarly, we define a maximum attachment probability $P_{max}$, which represents the affinity of the pilus tip to binding sites provided by the EPS, and corresponds to the situation where $K = 1$.

If the attachment point in question lies on the body of another particle, the probability of particle-particle attachment ($P_{b}$) is invoked first, if no attachment results the surface attachment probability ($P_{s}$) is evaluated (Fig. \ref{FigureS1}). 

\begin{table}
\newcommand{\mline}[2][c]{%
  \begin{tabular}[#1]{@{}c@{}}#2\end{tabular}}
\caption{parameters} % title of Table
\centering % used for centering table
\begin{tabular}{c c c} % centered columns (4 columns)
\hline\hline %inserts double horizontal lines
parameter & effect & value(s) used \\ [1ex] % inserts table
&\underline{rod mechanics}&\\[1ex]
%heading
\hline % inserts single horizontal line
$w$ & rod width & $1$ \\ % inserting body of the table
$l_{min}$ & minimum rod length & $3w$  \\
$l_{max}$ & maximum rod length & $7w$  \\
$F_{r}$ & repulsive force constant & $1$\\ 
$\mu$ & friction coefficient & $1$  \\[1ex]
 % [1ex] adds vertical space

\hline\hline %inserts double line
&\underline{motility}&\\[1ex]
%parameter & effect & value(s) used \\ [0.5ex]
\hline
$|\vec{F}_{p}|$ & pilus retraction force & $1.5$  \\
$r_{pili}$ & length of pilus & $l_{min}$  \\
$\phi$ & angular range of pilus & $0.5\pi$\\
$t_{ret}$ & \mline{mean retraction time} & $5s$\\
$t_{rev}$ & mean reversal period & $1000s$\\ 
$\sigma_{rev}$ & st. dev. $t_{rev}$& $t_{rev} / 5$ \\

\hline\hline %inserts double line
&\underline{furrowing}&\\[1ex]
%parameter & effect & value(s) used \\ [0.5ex]
\hline
$\gamma$ & stiffness of substratum & $[0.001, 1.6]$  \\[1ex]
$k_{U} = \gamma k_s $ & deformation rate & $0.05$  \\
$\beta_{U} = \gamma \beta_s $ & \mline{restitution rate} & $2.5\times10^{-4}$ \\

\hline\hline %inserts double line
&\underline{EPS trails}&\\[1ex]
%parameter & effect & value(s) used \\ [0.5ex]
\hline
$P_{min}$ & \mline{minimum pilus \\ attachment probability} & $0.1$  \\
$P_{max}$ & \mline{maximum pilus \\ attachment probability} & $0.3$ \\
$k_{p}$ & EPS deposition rate & $0.1$\\
$\beta_{p}$ & EPS degradation rate & $5\times10^{-4}$\\

\hline\hline
&\underline{implementation}&\\[1ex]
%parameter & effect & value(s) used \\ [0.5ex]
\hline
$\Delta x$ & grid resolution & $w/4$ \\
$C_{max}$ & max. tracer count per pixel & $1\times [\Delta x]^2$\\
$L$ & simulation box dimension & $160$ \\ 
$N$ & number of particles & $1000$  \\
$t_{f}$ & simulation duration & $5.0\times10^{4}s^{*}$  \\
$t_{rec}$ & data acquisition interval & $20s$ \\
$F_{max}$ & max. allowed repulsion force & $10$\\
$\Delta t_{max}$ & maximum discrete time step & $0.1 s$\\
$\Delta t_{min}$ & minimum discrete time step & $0.005 s$\\ 
\hline\hline

\end{tabular}
\begin{tabular}{c}
*for $\gamma = [0.9, 1, 1.1, 1.3]$, $t_{f} = 1.0\times10^{5}$\\
\end{tabular}
\end{table}
\label{table1}
\label{tableS1}

% parametrization
\subsection{parameterization}

While precise estimates do not exist for many of the model input parameters, it has been documented that during unobstructed twitching motility {\it P. aeruginosa} move discontinuously at approximately 0.3 - 0.5 $ {\mu}$m$/s$ in excursions of approximately 5 ${\mu}$m, and pili can reach through an arc of at least $~0.5\pi$ radians \cite{skerker2001direct}. The average polarity reversal rate and its distribution are important parameters with respect to collective motion \cite{starruss2012pattern} and could be subject to a biochemical regulation mechanism \cite{mattick2002type,shi1996cell}. This regulation process is not well-characterized, and may be unique to specific species, strains and conditions. Since there are no long-range chemical gradients in our simulations, we make the minimal assumption that bacteria reverse polarity several times during a division cycle, which can range from about 30~min to 200~min depending on conditions (see for example \cite{yang2008situ,gottenbos2000initial}). Visual inspection of the microscopy data in \cite{gloag2013self} indicates that bacteria range from about 3 - 7 ${\mu}$m in length and are about 1 ${\mu}$m in girth.  

Based on the observations and limitations discussed above, the following parameters can be roughly established: $t_{rev}$, $\sigma_{rev}$, $l_{max}$, $l_{min}$, $r_{pili}$, $\phi$, and $w$.  The observed unobstructed movement rate of 0.5 $\mu$m/$s$ is established by balancing $t_{ret}$, $|\vec{F_{p}}|$, and $\mu$. Biological parameters that must be estimated are: $k_{p}$, $P_{max}$, $P_{b}$, and $F_{r}$. The repulsion force parameter $F_{r}$ is simply set to a value that prevents overlap of rod centerlines during collisions. We set the EPS deposition rate $k_{p}$, to a value that allows the steady state EPS concentration to be reached in the time that a particle passes over an area. 

We set the maximum surface attachment probability $P_{max} = 0.3$, a value that approximately reconciles the maximum observed long-time movement rate observed for {\it{P. aeruginosa}} in the interstitial environment of 0.2 $\mu \text{m}/s$ and the maximum single-retraction velocity of 0.5 $\mu \text{m} /s$. We set the affinity of the pilus tip to the bacterial cell surface $P_{b} = 0.25$, slightly lower than the maximum possible EPS attachment probability. This decision was based on the idea that attachment to the cell surface would likely be mediated by EPS materials which would have to bind to the pilus tip as well as the cell surface for movement to result from retraction.  We set $P_{b}$ higher than $P_{min}$ based on the assumption that attachment to a cell surface would be favored relative to the native surface. 

The experiments we wish to simulate (systematic alteration of substratum conditions) do not alter the bacteria. Therefore, their physical and biological properties were fixed in the model.
 
With biological parameters fixed, those relating to the environment can be systematically tuned and the results compared to experimental observations \cite{gloag2013self,semmler1999re}. The environmental parameters are: $\gamma$, $k_{s}$, $\beta_{s}$, $\mu$, $P_{min}$, and $\beta_{p}$. 

The minimum attachment probability, $P_{min}$ corresponds to the affinity of the pilus tip to either the substratum or the coverslip surface. In order for the stigmergic effects of EPS excretion to manifest, $P_{min}$ must be lower than $P_{max}$ so that spatial variations in binding probability can bias movement. 

The time scale associated with topographical dynamics in the substratum is defined by $\beta_{s}$ and $k_{s}$, which are important parameters with respect to pattern formation. If $\beta_{s}$ is too high, the substratum will immediately re-form after passage of a particle and individuals or clusters will become isolated from each other as the substratum closes behind them. The rate ($k_{s}$) at which particles deform the substratum determines how long it takes for a stationary particle to generate a topographical gradient around itself. Therefore, (for fixed $\gamma$, $\mu$, and $|\vec{F_p}|$), $k_{s}$, $t_{ret}$, and $P_{min}$ will determine whether or not movement of individuals is observed.   

%initialization

\subsection{Initialization}

In each simulation, 1000 particles were placed in random positions throughout a square space (side length $L = 160$ $\mu$m) with periodic boundary conditions. Particles were assigned orientations ($\theta$), and lengths ($l$) from uniform distributions $[0, 2 \pi)$ and $[l_{min}, l_{max}]$, respectively. Reversal clocks were initially randomized by assigning each particle a reversal period from the Gaussian distribution described by $t_{rev}$ and $\sigma_{rev}$, and allowing the values to count down and reset according to the rules of the model for a time period of $10 t_{rev}$ before starting the simulation. Pili are assumed to be initially unattached to the surface, with the countdown to the first attachment attempt selected from the uniform distribution $[0, t_{ret}]$.

\section{results}

%introduction
The only parameter that we varied in order to cause the changes in collective behavior discussed below is the stiffness coefficient ($\gamma$) which governs the resistance experienced by the motile bacteria, the rate at which they deform the local topography, and the rate at which the furrows refill. Altering $\gamma$ is akin to changing the concentration of monomeric precursor when preparing the substratum (eg: agar or gellan gum), or altering the concentration of stabilizing di-valent cations in the mixture. 

\begin{figure}[b!]
\centering
\includegraphics[width=0.9\linewidth]{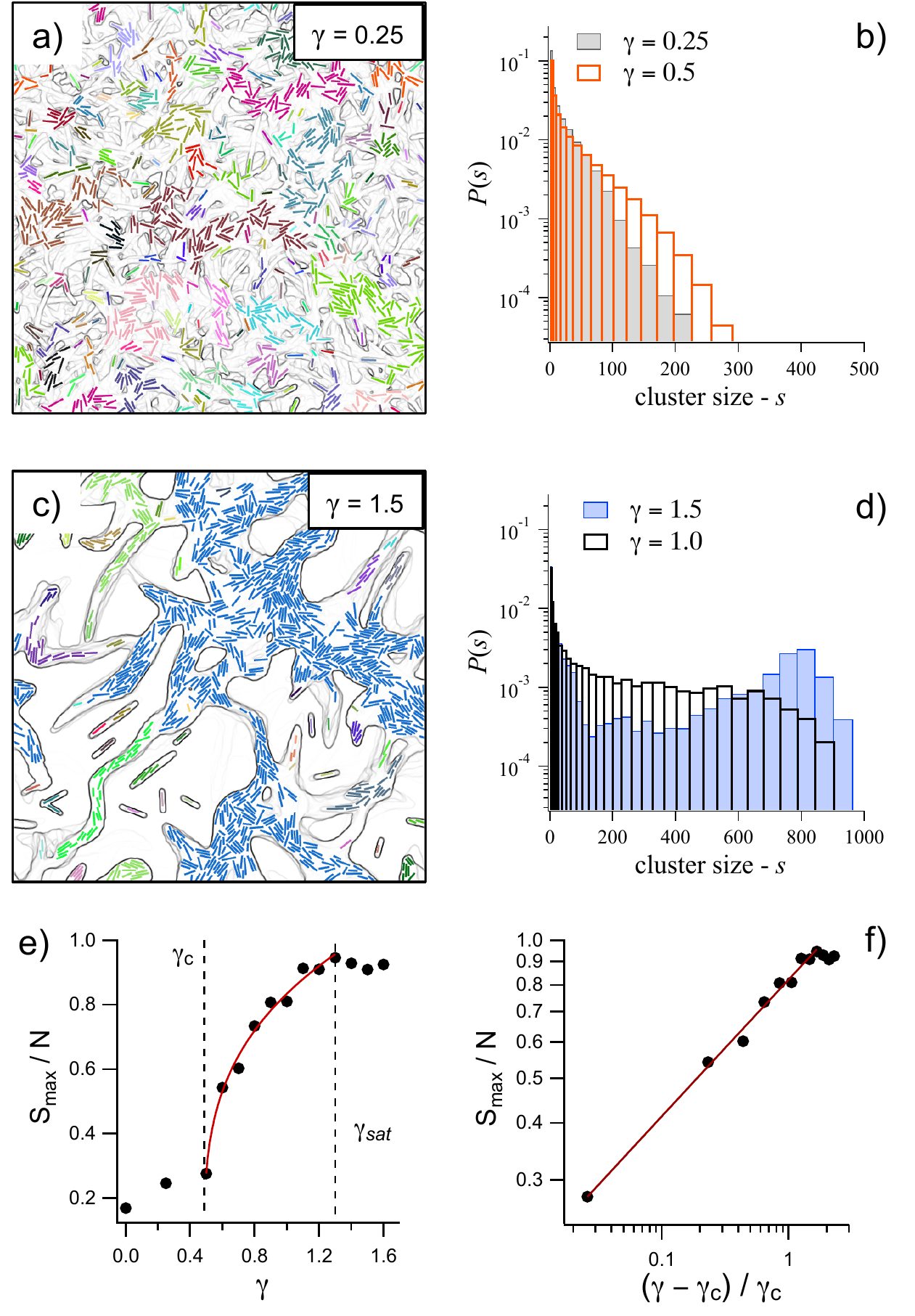}
\caption{Clustering induced by a change in substratum stiffness, $\gamma$. (a, c) Snapshots of the colony illustrating the the influence of $\gamma$ on clustering in the steady state. Bacteria in each cluster are given the same (arbitrary) color and the furrow edges are shown in gray. (b, d) The cluster size distribution, $P(s)$, transitions from an exponential decay to a bi-modal distribution as $\gamma$ is increased from 0.25 (low substratum stiffness) to 1.5 (high substratum stiffness). (e) Maximum cluster size $S_{max}$ (normalized to the system size of $N$ bacteria) versus $\gamma$, revealing a clear, critical value $\gamma_c$ that marks the onset of stable trail formation. The red line indicates the transition region between unpercolated $(\gamma\leq\gamma_c)$ and percolated $(\gamma\geq\gamma_{sat})$ states. (f) The maximum cluster size obeys a power law as a function of $\gamma$ within the transition region, which we confirm by rescaling $\gamma$ to $\gamma_c$.}
\label{CSD_PT}
\end{figure}

%qualitative CSD ``percolation''
We start by characterizing clustering and connectivity in the simulated biofilm for systematically increasing values of $\gamma$. The steady state cluster size distribution (CSD), defined as the probability density $P$ of finding an individual within a cluster of size $s$, is affected by the physical properties of the substratum, as is demonstrated by the plots of $P(s)$ versus $\gamma$ shown in Fig. \ref{CSD_PT}. At low values of $\gamma$, the bacteria experience little resistance to movement through the substratum, no clear trail systems are visible in snapshots of the colony in the steady state (Fig. \ref{CSD_PT}a, Supplemental Movie S1 \cite{Supp}), and the CSD decays exponentially (Fig. \ref{CSD_PT}b). As $\gamma$ is increased, the bacteria form networks of interconnected trails (Fig. \ref{CSD_PT}c, Supplemental Movie S2 \cite{Supp}) and the CSD transitions to a bimodal distribution (Fig. \ref{CSD_PT}d) \cite{Supp}. The formation of large clusters that approach the system size (of 1000 bacteria) is observed when the resistance to motion approaches the force exerted by individuals attempting to move through the substratum. We refer to this process as {\it percolation}, because it indicates an approach to global connectivity between most of the individuals in the system. Though a finite-size study is outside the scope of the present work, it is possible that for the particle density we used in these simulations, percolation would not occur in much larger systems, and the maximum cluster size may instead converge to some value smaller than the system size. Percolation theory suggests that this tendency would depend sensitively on the particle density.

%transition to percolation
To quantify the transition from equilibrium clustering to percolation, we plot the maximum cluster size $S_{max}$ reached for each value of $\gamma$, and find distinct scaling behaviors below and above a critical value $\gamma_c \approx 0.49$ (Fig. \ref{CSD_PT}e), which marks the onset of the transition to the percolated state. The transition region $(\gamma_c\geq\gamma\geq\gamma_{sat})$ is highlighted by the red line in Fig. \ref{CSD_PT}e. The maximum cluster size distribution obeys a power law within the transition region (Fig. \ref{CSD_PT}f), and the system resides in stable unpercolated and percolated states below $(\gamma\leq\gamma_c)$ and beyond $(\gamma\geq\gamma_{sat})$ the transition region. The change in clustering behavior around $\gamma_c$ is extremely abrupt, and a cautionary sign for experimentalists -- a subtle change in a physical parameter such as stiffness around a critical value can produce a dramatic change in colony behavior that may be caused purely by unintended changes in a poorly-controlled parameter. 

%peruani's transition
The change in CSD from an exponential to a bi-modal distribution seen in Fig. \ref{CSD_PT}b,d is not unique to our study. A similar transition has been observed in prior simulation studies of self-propelled rods, where it indicates a change from a state of uncorrelated movement to one of ordered collective motion \cite{peruani2006nonequilibrium}. Experiments with the bacterium {\it{Myxococcus xanthus}} have shown similar behaviors, with long-tailed or bi-modal CSDs occurring below and above a critical cell density \cite{starruss2012pattern,peruani2012collective}. In these studies, a critical density marks the transition from uncorrelated movement to collective motion, which is facilitated by high degrees of orientational alignment between individuals in contact. More broadly, density and orientational order/disorder are fundamentally important in systems of moving agents, and constitute the two most basic control parameters in minimal models of collective motion \cite{vicsek1995novel}. 
%transition to orientational coherence and local density

We therefore proceed by analyzing changes in orientational coherence (ie: degree of alignment between neighboring bacteria) and local density associated with the changes in clustering and percolation seen in Fig. \ref{CSD_PT}. 

\begin{figure*}[t!]
\centering
{\includegraphics[width=0.9\textwidth]{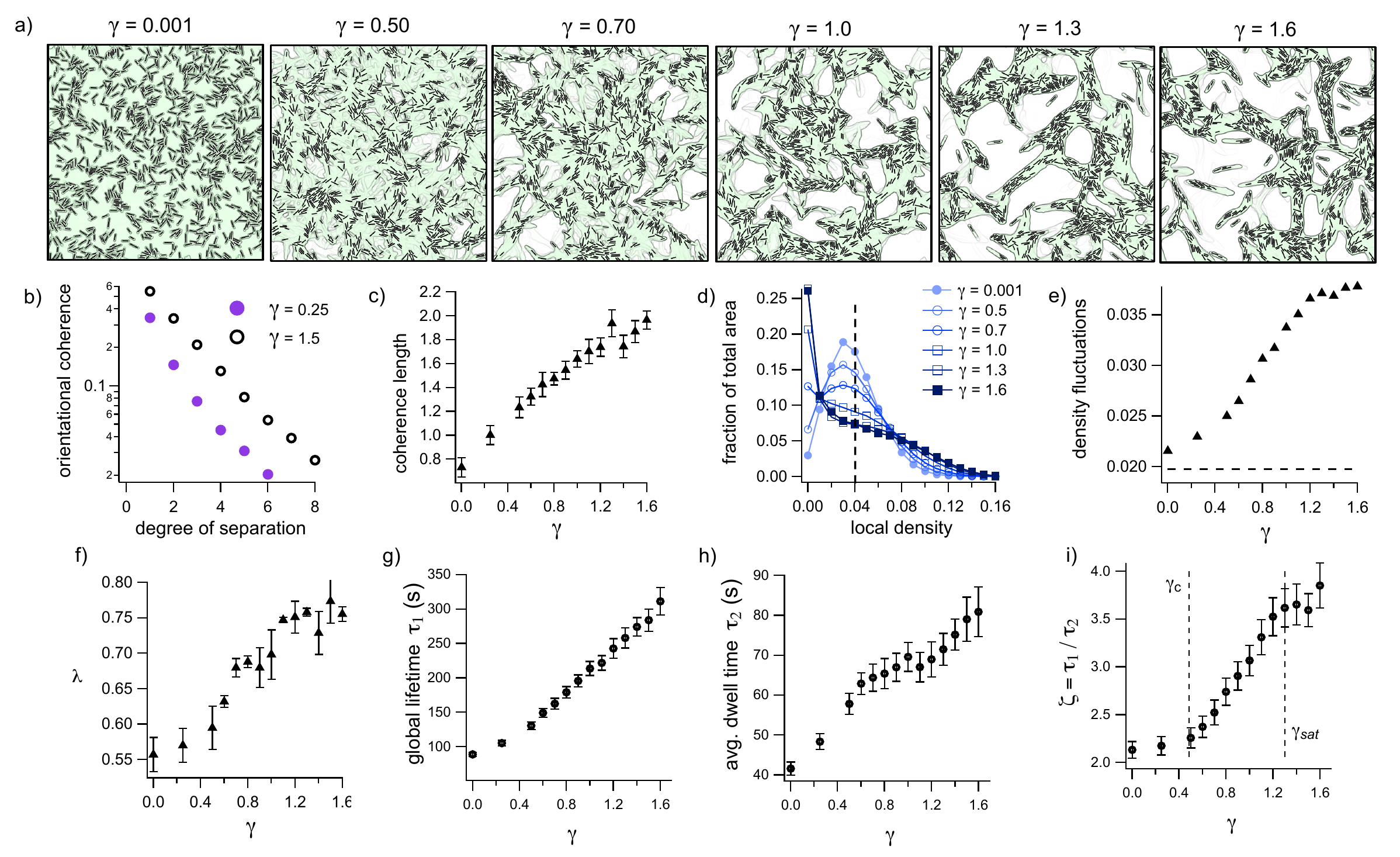}}
\caption{{Steady-state properties of the biofilm versus substratum stiffness $\gamma$.} (a) Snapshots of the colony superimposed on plots of EPS concentration (shown in green). The local degree of alignment increases with $\gamma$. It manifests as an increase in orientational coherence (b) and coherence length versus $\gamma$ (c). Clusters form throughout the colony as $\gamma$ is increased, causing broadening of the spatial density distribution (d) and an increase in local density fluctuations (e), which correspond to non-equilibrium number fluctuations (f). The global configuration (ie: colony morphology) persistence time (g) and the individual bacterium dwell time (h) both increase monotonically with $\gamma$. In the percolated state $(\gamma\geq\gamma_{sat})$, the colony morphology is more stable than in the unpercolated state $(\gamma\leq\gamma_c)$, relative to the movement rate of the individuals composing it. This indicates percolation in conjunction with path-following (i). Error bars correspond to the standard deviation in time (g, i), or the 95\% confidence interval of the fitting coefficient (c, f).}
\label{OC_dD_z}
\end{figure*}

%orientational coherence
Qualitatively, the snapshots in Fig. \ref{OC_dD_z}a indicate the existence of two distinct states at low and high values of $\gamma$, and that the local density and degree of alignment between neighboring bacteria increase with $\gamma$, as the system transitions from the unpercolated to the percolated state. To quantify local alignment, we compute the orientational coherence ($\phi$, defined by Eq. \ref{phi_vs_dos} see Methods) between bacteria versus the degree of separation (DoS). For two bacteria in contact DoS $= 1$, for second nearest neighbors DoS $= 2$ and so on. For bacteria with parallel orientations, $\phi = 1$ (fully correlated), and for bacteria with perpendicular orientations $\phi = -1$ (anti-correlated). The steady-state time average of the ensemble mean of $\phi$ decays exponentially as DoS increases (Fig. \ref{OC_dD_z}b), and the reciprocal of the corresponding decay constant (the coherence length) quantifies local orientational order. Fig. \ref{OC_dD_z}c shows that the coherence length increases continuously with $\gamma$, and does not show a clear demarcation between the unpercolated and percolated states. However, the trend is clear, and confirms that local alignment does indeed increase with $\gamma$, in agreement with qualitative inspection of the colony snapshots in Fig. \ref{OC_dD_z}a. We note that the snapshots in Fig. \ref{OC_dD_z}a indicate alignment correlates with high EPS concentration, even though EPS does not cause explicit alignment of neighboring bacteria in our model. Instead, alignment occurs through collisions of rod-shaped bacteria which are confined within the furrows. 

%density distribution
The local density distribution also changes dramatically with $\gamma$ (Fig. \ref{OC_dD_z}d). At low values of $\gamma$, the local density distribution peaks near the global density $(N/L^2)$ of 0.04 (where $N$ and $L$ are the total number of bacteria and the length of the simulation box respectively), as expected for a homogeneous system. Conversely, at high values of $\gamma$, the distribution peaks at 0 (corresponding to empty regions between the clusters seen in Fig. \ref{OC_dD_z}a), and spreads out to high densities encountered within the clusters. The spreading corresponds to an increase in local density fluctuations (defined as the standard deviation of the local density) which grow systematically from the equilibrium value (indicated by a dashed line in Fig. \ref{OC_dD_z}e) as $\gamma$ increases. Figures \ref{OC_dD_z}d,e represent density fluctuations for square subspaces of side length $10\mu$m. However, it is instructive to examine the density fluctuations as a function of subspace area (Fig. \ref{OC_dD_z}f). Below $\gamma_c$, particle number fluctuations scale as $\Delta n \propto \langle n \rangle ^ {0.5}$ (where $\langle n \rangle = N [l/ L]^{-2}$ is the average number of bacteria in a local area of area $l^2$), as expected for an equilibrium system. Above $\gamma_c$ non-equilibrium giant number fluctuations (GNF) are observed and $\Delta n \propto \langle n \rangle ^ {\lambda}$, with $\lambda > 0.5$ (see Methods for a detailed description of our local density calculations). The increase in density fluctuations above $\gamma_c$ is important with respect to bacterial fitness because it is the high-density tail of the density distribution that corresponds to tightly packed areas where the bacteria will benefit from cooperative processes \cite{ratzke2016self}, and are more likely to withstand perturbation.

%zeta
To investigate the balance between cluster stability and individual mobility, we introduce a novel metric: the configurational correlation, $\zeta$, defined as the ratio of the persistence time $(\tau_1)$ of a global density distribution to the dwell time $(\tau_2)$ of the constituent individuals. It is a quantitative measure of the rate at which the colony `morphology' seen in each snapshot of Fig. \ref{OC_dD_z}a evolves relative to the rate at which the bacteria within it are moving. A complete description of how we computed $\tau_1$, $\tau_2$ and $\zeta$ is provided in the Methods section.

Plots of $\tau_1$ (Fig. \ref{OC_dD_z}g) and $\tau_2$ (Fig. \ref{OC_dD_z}h) versus $\gamma$ show, unsurprisingly, that both the global persistence time and and the individual dwell time increase monotonically with substratum stiffness. However, $\zeta(\gamma)$ is approximately constant in the unpercolated and percolated states, and increases continuously with $\gamma$ in the transition region between the two states (Fig. \ref{OC_dD_z}i). The transition region, $\gamma_c\leq\gamma\leq\gamma_{sat}$, demarcates a state characterized by dispersed, disordered movement $(\gamma\leq\gamma_c)$, from that of a structured collective morphology with high local density and connectivity$(\gamma\geq\gamma_{sat})$. Thus, there exist two ranges of substratum stiffness, corresponding to the two stable states of the colony within which the respective modes of collective movement are insensitive to small changes in $\gamma$. The stable states are separated by a rapidly-varying transition region characterized by a high degree of sensitivity to $\gamma$. 

For $\gamma \leq \gamma_{c}$, the self-propulsion force of individual bacteria is significant relative to the opposing force exerted by the substratum. Hence, individuals move freely, forming clusters only when random collisions occur (Fig. \ref{model}b), and disperse rapidly due to the stochastic nature of movement driven by Type IV Pili (see Fig.\ref{model}c and the model description). In this sub-critical regime, path following and stable clustering do not occur because the two stigmergy mechanisms are ineffective -- the substratum does not produce robust topographic paths, and the EPS concentration rapidly becomes uniform throughout the system, as is seen in Fig. \ref{OC_dD_z}a where the EPS concentration is depicted in green. Hence, movement bias caused by the environment is negligible, and stigmergy does not occur. 

As $\gamma$ increases, the substratum increasingly restricts the motion of individuals, and the motility bias induced by EPS causes separated individuals to form clusters. Polarity reversals give rise to cluster elongation and allow bacteria to semi-periodically retrace their steps, reinforcing trails that eventually merge, forming the observed network of interconnected channels. 

%discussion
\section{discussion}
Having characterized the transition between the two modes of collective behavior, we first discuss the essential features of this process and the underlying mechanisms. We then discuss how these results help us understand the interplay between biofilm fitness and morphogenesis phenomena reported in experimental studies.

%\subsection{assumptions and predictions}
The primary aim of this study was to elucidate the nature of biofilm pattern formation based on the assumptions that the constituent bacteria furrow through the substratum, and excrete EPS which can form trails that bias their motion. We note that, whilst the model implements two forms of stigmergy (due to furrowing and EPS following), EPS alone can give rise to trail and pattern formation even in the absence of furrowing (eg. when $\gamma = 0$). However, this is true only in a narrow subset of the parameter space where the EPS degradation and deposition rates are fine-tuned and the pilus binding affinity to the bare surface is close to zero -- conditions that are not justified given that the bacteria are known to form furrows in the systems of interest \cite{gloag2013self}, and that T4P are known to bind readily to a large variety of biological and abiotic surfaces \cite{giltner2006pseudomonas}. 

Additionally, low-concentration agar is known to favor `swarming' or `swimming' motility through the expression of flagella (see for example \cite{kearns2010field,tremblay2008improving}), mechanisms distinct from the one we study here. Our simulations of twitching motility are expected to apply in situations where substratum stiffness is high enough to promote the expression of T4P, which makes the substratum stiffness an intrinsic parameter in any realistic scenario involving this type of motility.

While bi-modal clustering in systems of self-propelled rods has been attributed to simple steric interactions between particles \cite{peruani2006nonequilibrium}, such a transition is not evident in our model, which is based on stochastic, apolar movement of bacteria as documented by Skerker et al. \cite{skerker2001direct}. Indeed, below $\gamma_c$ the system demonstrates equilibrium behavior, even though the cell density and mean aspect ratio are well within the parameter space corresponding to non-equilibrium clustering as reported by Peruani  et al. \cite{peruani2006nonequilibrium}. The absence of this transition is due to the stochastic nature of particle movement, in particular the discontinuous application of the pilus retraction force, and random polarity reversal \cite{Supp}. However, trail following due to substratum deformation effectively facilitates non-equilibrium structure formation, despite the stochastic nature of bacterial movement. Importantly, our results indicate that EPS deposition alone is not sufficient to achieve trail formation.

This result both complements and contrasts that of Balagam and Igoshin \cite{balagam2015mechanism}, who recently simulated trail-following in surface-motile bacteria to explain pattern formation in  systems where the constituent individuals spontaneously reverse direction. While they observed that trail-following does indeed facilitate non-equilibrium clustering, the authors did not suggest a specific mechanism behind trail-following phenomena but invoked the idea of EPS trail following to explain the corresponding experimental observations. Their implementation of trail sensing involved a monte-carlo sampling of the area near the forward pole of each cell to determine the trail direction. In our model such a process is analagous to the random sampling of the area accessible to pili. However, because pilus retractions pull the cell a distance on the order of the cell length \cite{skerker2001direct}, errors in spatial sampling lead to ineffective EPS trail following. Our results suggest that while EPS excretion plays an important role in morphogenesis by enhancing the movement rate within trails, these trails cannot form without the mechanism suggested by Gloag et al. \cite{gloag2013self}, whereby the bacteria actively remodel the topography of their environment. Without this effect, trail-following by the preferential binding of T4P to EPS is not robust.

For the reasons discussed above, we expect our model to exhibit a robust behavioral dependence on $\gamma$ similar to the one we report here for any set of parameters for which the EPS does not facilitate path following in the absence of a physical resistance to motion. The specific range of $\gamma$ over which this transition occurs experimentally will be a function of the forces exerted by the bacteria. Therefore, we would not expect to see the emergence of robust trail networks in experiments where the bacteria move through a medium that is either very soft (negligible $\gamma$) or very rigid (negligible $k_s$), unless the pilus binding affinity is close to zero in the absence of EPS.

Our insights into the role of $\gamma$ are highly relevant to experimental studies because, whilst EPS  following is difficult to manipulate experimentally, the material properties of the culture medium are relatively easy to control and can have an enormous influence on behavior. This indicates that unpredictable behavior of bacterial biofilms in the laboratory may result from subtle differences in culture preparation, and motivates novel studies of the behavioral response of bacteria to systematic manipulation of material properties.

An example of this strategy was recently reported by Ratzke et al., \cite{ratzke2016self} who elegantly demonstrated the trade-off between fitness associated with low and high-resistance environments for micro-colonies of {\it B. subtilis} (a motile, rod-shaped bacterium) suspended in solidified agar. In low-density media, bacteria were able to move freely into the medium, expanding rapidly and proliferating quickly in high-nutrient conditions, while their counterparts grown in high-nutrient, high-density agar were not able to expand rapidly and formed dense patches. In low-nutrient conditions, however, the advantage was given to the high-density, slow-moving structures which were able to survive due to a cooperative metabolism process that was only possible in dense patches. These findings suggest a mutual exclusivity separating high-mobility and high-density structures. 

Our results show how stigmergic processes can combine these two fitness advantages in the high-density condition. Cooperative movement processes and trail-formation produce a system in which the individuals can move effectively within a densely connected structure that maintains the benefits of the biofilm without inhibiting spatial proliferation.  

\section{Conclusion}
We have demonstrated how slight alterations in environmental conditions can lead to dramatic changes in collective motion patterns of bacteria, even when all biological parameters are fixed. The observed behavioral transition is facilitated by an interplay between two stigmergic processes. The first is biological: the bacteria actively store spatial information in the local environment in the form of excreted substances. The other is physical: spatial information is stored due to deformation of the material through which the bacteria move. The physical mechanism offers an opportunity to understand and control the behavior of bacteria without modifying their biology. Our results emphasize the essential role of agent-environment interactions in models of emergence in multi-agent systems, particularly when the environment is not static, and is modified by the activity of the agents moving within it.

\section{methods}

\subsection{simulations}

\begin{figure}[h!]
\includegraphics[width=0.45\textwidth]{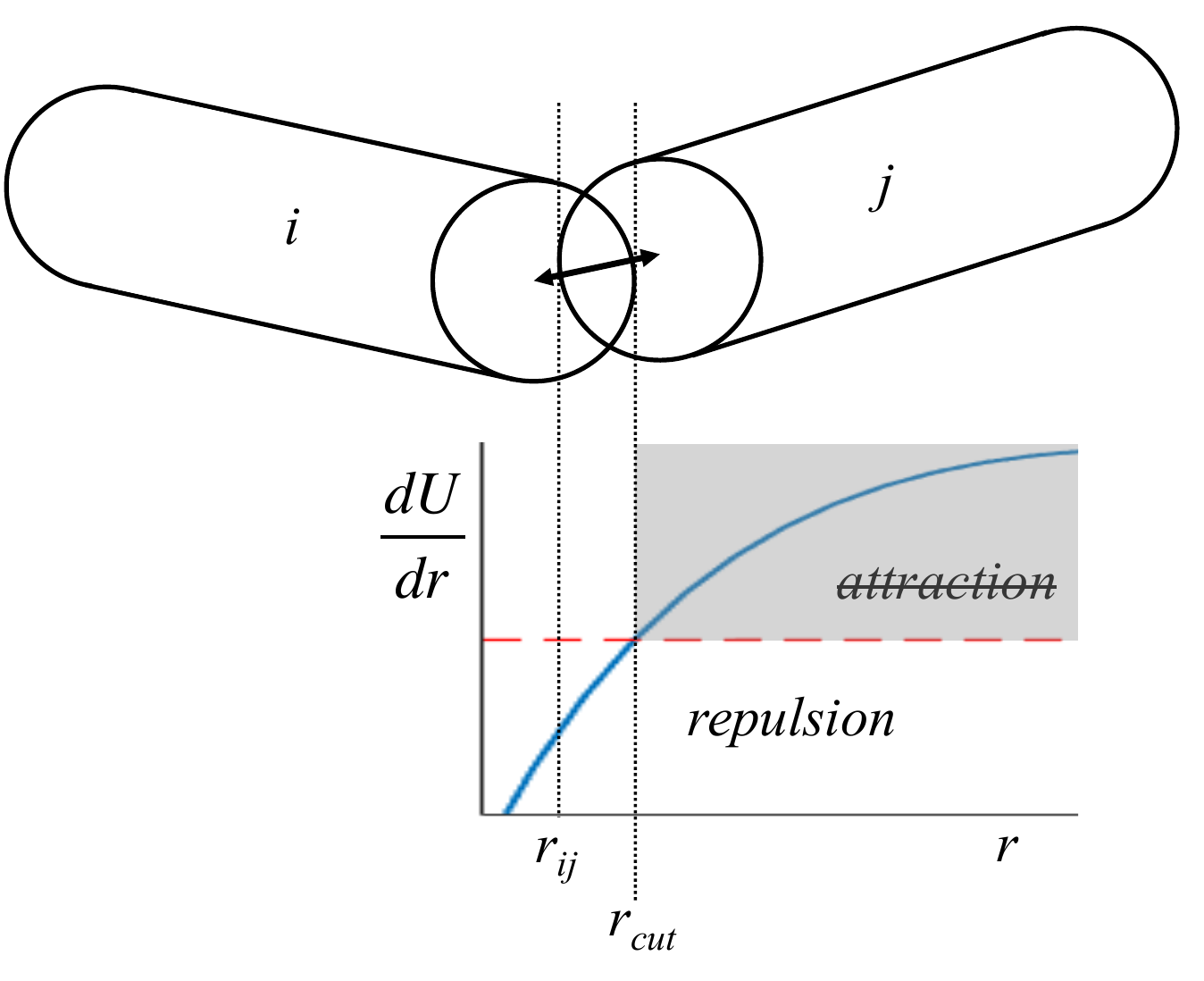}

\caption{Binary interactions between individuals: If the distance between two particles is shorter than the rod width, repulsion occurs between them. The inset illustrates the interaction force defined by a Lennard-Jones potential where $r_{eq} = r_{cut} = w$ (the attractive part of the potential is ignored).}
\label{FigureS2}
\end{figure}

We used the Vega-Lago algorithm \cite{vega1994fast} in to find distances between particle backbones. We note that the FORTRAN code provided in \cite{vega1994fast} contains an error that causes the algorithm to fail for parallel rods of different length, which we corrected in the current implementation. We also used a cell-sorting algorithm (side length $l_{max} + w$) and neighbor lists to efficiently find the distances between individuals. $|\vec{F}_{ij}|$ is calculated from the repulsive component of a Lennard-Jones potential (Fig. \ref{FigureS2}). The scaling of the repulsive force is given by $F_{r}$ which is the maximum value reached by the attractive portion of the first derivative of the potential. Though the attractive part is ignored, this value adjusts the scaling of the repulsive region, so that a higher value of $F_{r}$ yields stronger repulsion.

We used Eulerian integration to calculate changes in particle position. The time step $\Delta t$ is continuously adjusted between $\Delta t_{max} = 0.1$~s and $\Delta t_{min} = 0.005$~s so that changes in particle positions do not exceed a specified maximum. To avoid numerical artifacts when $\Delta t = \Delta t_{min}$, a maximum repulsive force threshold ($F_{max}$) is implemented. As long as $F_{max}$ is set to values much higher than typical interaction forces, over-compression of the particles does not result in overlap between particle centerlines (a physically impossible arrangement). The main reason for implementing this force threshold is to allow overlaps from the initially random configuration to resolve without displacements exceeding those that would result from typical particle-particle collisions. 

\subsection{data analysis}

The principle output of any self-propelled rod simulation is a set of configurations described by the particle positions (Cartesian coordinates), orientations, and lengths. To analyze simulation output, we constructed a sparse adjacency matrix using a cutoff nearest-neighbor distance $r_n$ = $1.5w$ larger than the physical repulsion distance but small enough to exclude second-nearest neighbors. 

The cluster size probability density (CSD) represents the probability that a particle selected randomly in the steady state will belong to a cluster of a given size. CSD values are computed from an adjacency matrix: neighbors of increasing degree are identified until the unique list of particles within the cluster does not change with further iterations. The cluster size is the number of particles identified in this way. Histograms of the cluster size distribution were plotted by binning the cluster size values for each particle into bins of increasing size ($ s = 1^2, 2^2, 3^2 ...  32^2$), averaging over the steady-state time period, and normalizing by bin width to reflect the probability ($P(s)$) of randomly identifying a particle within the size range defined by each bin. 

To calculate the average nematic orientational coherence value for the ensemble as a function of DoS between particles we used the following procedure: for each particle, the orientations of neighbors at increasing DoS are compared to that of the particle in question and averaged via the orientational coherence function:

\begin{equation}
\phi_{i}(DoS) = \frac{\sum\limits_{j = 1}^{n} 2[\cos(\theta_{i} - \theta_{j})]^{2} - 1}{n(DoS)}\,,
\label{phi_vs_dos}
\end{equation}

and the ensemble average, $\langle\phi\rangle_{N}$ vs DoS over the entire system is calculated. 

We computed density distributions and their fluctuations by dividing the simulation space (L = 160) into square subspaces $\varsigma$ of side length $l$, and counting the number of particles with centroids located in each subspace, to compute local densities $\rho(\varsigma, t)$ in each time step. The values in Figure \ref{OC_dD_z}d represent the fractions of the total area containing the indicated particle densities for subspace area $a = 100\mu$m$^2$. These are calculated by counting the number $q$ of subspaces containing the indicated density $\rho(\varsigma)$ and computing the time average of $q a / L^2$ for the steady state. This results in a single histogram of local density for the entire steady state. The fluctuation magnitude (Fig.\ref{OC_dD_z}e) is simply the standard deviation of these discrete local density values. 

The configurational correlation $\zeta$ is the ratio of two different configurational decay times. The first, $\tau_{1}$, is a measure of the time it takes for a global configuration to cease resembling itself. It is computed as follows: the simulation box is divided into a square mesh with resolution set equal to the minimum particle length.  At time $t_{i}$ each cell in the mesh is assigned a 1 if occupied by at least one particle, or 0 otherwise, we refer to this as the structural configuration matrix ($S(t_{i})$). 

We take the sum over the elements of $S(t_{i})$ to find $z(t_{i})$. $S(t_{i})$ is then multiplied element-wise with the subsequent configuration at $t = t + dt$ and the result is $S(t_{i} + dt)$ this process is repeated for subsequent configurations and we find $R(t - t_{i})$, the normalized sum over all elements of $S(t)$, so that $R(t - t_{i}) = z(t-t_{i}) / z(t_{i})$ approaches zero as $t - t_{i}$ increases. Fitting $R(t - t_{i})$ to a single exponential decay gives $R(t - t_{i}) = e^{[-1/\tau_{1}][t - t_{i}]}$. 

The other type of configurational decay constant, $\tau_{2}$, is calculated similarly: at $t_{i}$, each particle is assigned an index corresponding to its grid position, and the correlation value, $\kappa(t_{i}) = 1$. If a particle's index changes in subsequent configurations the correlation value $\kappa(t - t_{i})$ decreases by $1/N$, the decay is cumulative in time, ie: $\kappa(t-t_{i})$ does not recover if a particle goes back to its location at $t_{i}$. The decay is fit to an exponential model so that $\kappa(t-t_{i}) = e^{[-1/\tau_{2}][t-t_{i}]}$. For a given simulation in the steady state $\zeta = \tau_1 / \tau_2$ is a constant that fluctuates about a defined mean.

To ensure that the system had reached steady-state behavior, time-dependent cluster size distributions were examined to establish whether the system's CSD had reached a stable trend. All simulations had reached steady-state CSDs by $t = t_{f} - t_{f} / 4$ so the final quarter of each time series was used for all calculations. 

\section{acknowledgments}

CZ would like to acknowledge Matthew Arnold, Prabhakar Ranganathan, and Jared Cullen for useful discussions. This work was funded by FEI Company and the Australian Research Council (DP140102721 and CE110001018).

\section*{References}
\bibliography{latticing_refs}

\clearpage

\newcommand{\beginsupplement}{%
        \setcounter{table}{0}
        \renewcommand{\thetable}{S\arabic{table}}%
        \setcounter{figure}{0}
        \renewcommand{\thefigure}{S\arabic{figure}}%
        \setcounter{page}{1}
        \renewcommand{\thepage}{S\arabic{page}} 
     }

\section*{Supporting Information for\\`Emergent pattern formation in an interstitial biofilm'}
\centerline {by Zachreson \it et al.}
\hspace{1cm}
\beginsupplement

\subsection{supplemental movies}

Movie S1 corresponds to Fig. \ref{CSD_PT}a and demonstrates bacterial collective behavior when substratum stiffness $\gamma = 0.25$ (the disordered, equilibrium state). Bacteria are colored based on their orientation (red cells are aligned with the x axis, and blue cells to the y axis). Light-colored trails correspond to high EPS concentration and dark edges represent topographical gradients in the substratum. Time between frames is 1000s, frame rate is 10fps. 
\\\\
Movie S2 corresponds to Fig. \ref{CSD_PT}c and demonstrates bacterial collective behavior when substratum stiffness $\gamma = 1.5$ (the ordered state). Bacteria are colored based on their orientation (red cells are aligned with the x axis, and blue cells to the y axis). Light-colored trails correspond to high EPS concentration and dark edges represent topographical gradients in the substratum. Time between frames is 1000s, frame rate is 10fps. 

\subsection{cluster statistics in motility model variants}

\begin{figure*}[h]
\includegraphics[width=0.45\textwidth]{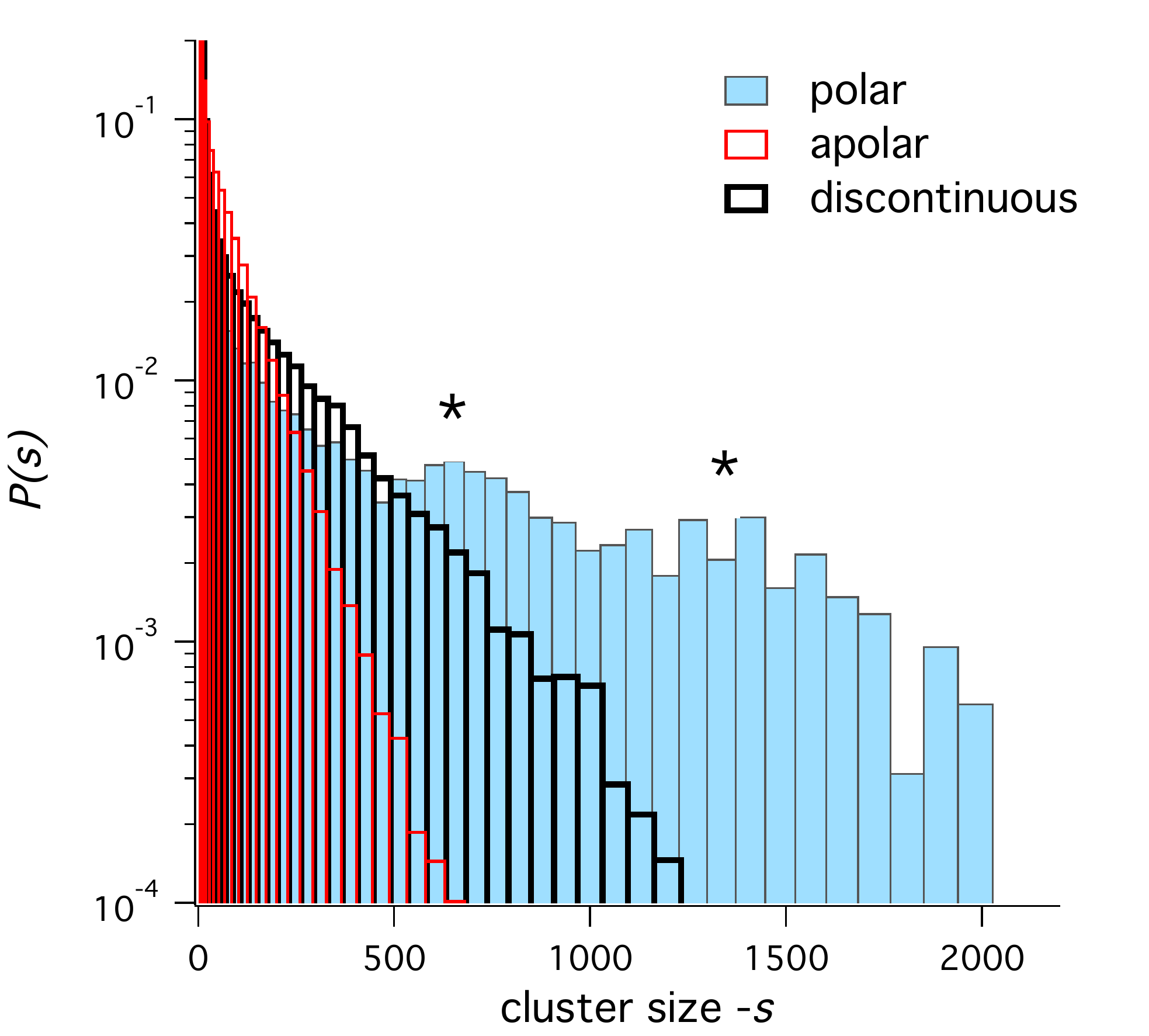}
\caption{Cluster statistics for simplified simulations of self-propelled rods. In the non-reversing case (blue bars), the distribution is multi-modal (asterisks (*) indicate non-equilibrium modes), in agreement with other models of self-propelled rods. However, this multi-modal cluster size distribution is not sustained if the particles either stochastically reverse direction (red), or move discontinusouly (black). }
\label{CSD_ideal}
\end{figure*}

To test the rules of our motility model against other, simpler models of self-propelled particles, we removed stigmergy from the model and tuned the parameters affecting particle movement to isolate various aspects of our system. Using the results of Peruani et al. \cite{peruani2006nonequilibrium} as a benchmark we tuned our model to mimic the behavior of systems of identical self-propelled rods that move in polar, continuous, deterministic paths. We achieved this by setting $l_{min} = l_{max} = 5$, $\phi_{ret} = 0$, and $P_s = 1$, and turning off direction reversals. In order to match Peruani et al's model we also had to reduce the propulsion force by a factor of 10 ($F_p = 0.15$) from the value used in the main text which was based on experimentally measured movement rates. The resulting cluster size distributions (Fig. \ref{CSD_ideal} blue bars) qualitatively resemble those reported by Peruani et al. (though our system is much larger, L = 240, N = 2986 for a coverage fraction of 0.3) wich likely accounts for the broadening of the non-equilibrium modes observed here.

We then made the following adjustments, individually, to the ideal model to investigate whether or not the elements of our motility model affect the tendency towards non-equilibrium clustering. We found that non-equilibrium clustering was robust to mixed aspect ratio ($l_{min} = 3$, $l_{max} = 7$), fluctuations in movement direction $\phi_{ret} = 0.5\pi$, and particle-particle attachment $P_b = 0.25$. However, stochastic direction reversals $T_{rev} = 1000s$ prevented clutering (Fig. \ref{CSD_ideal}, red bars). Additionally, implementing stochastic movement by setting $F_p = 10$ and $P_s = 0.1$ (for the same average movement velocity) also prevented the bi-modal clustering observed for the continuously moving case (Fig. \ref{CSD_ideal}, black bars). 

These tests indicate that we should not expect  collective motion for discontinuously moving rods that stochastically reverse direction. These attributes are essential to our model of {\it{Pseudomonas aeruginosa}}, which is designed based on observed behaviors \cite{maier2015bacteria,mattick2002type,skerker2001direct} and emphasizes the essential role of stigmergy in structure formation.

\end{document}